\documentclass{pasj01}
\usepackage{listings}
\usepackage{tabularx}
\usepackage{listings}
\usepackage{color}

\begin{document} 
\Received{}
\Accepted{}

\title{Acceleration of the tree method with SIMD instruction set}

\author{Tetsushi \textsc{Kodama}\altaffilmark{1}
}
\altaffiltext{1}{Department of Applied and Cognitive Informatics, Division of Mathematics and Informatics, Graduate School of Science and Engineering, Chiba University, 1-33, Yayoicho, Inage-ku, Chiba-shi, Chiba, Japan}
\email{adaa2540@chiba-u.jp}

\author{Tomoaki \textsc{Ishiyama},\altaffilmark{2}}
\altaffiltext{2}{Institute of Management and Information Technologies, Chiba University, 1-33, Yayoicho, Inage-ku, Chiba-shi, Chiba, Japan}
\email{ishiyama@chiba-u.jp}


\maketitle
\begin{abstract}
We have developed a highly-tuned software library that accelerates the
calculation of quadrupole terms in the Barnes-Hut tree code by use of
a SIMD instruction set on the x86 architecture, Advanced Vector
eXtensions 2 (AVX2). Our code is implemented as an extension of
Phantom-GRAPE software library 
that
significantly accelerates the calculation of monopole terms.  If the
same accuracy is required, the calculation of quadrupole terms can
accelerate the evaluation of forces than that of only monopole terms
because we can approximate gravitational forces from closer particles
by quadrupole moments than by only monopole moments.  Our
implementation can calculate gravitational forces about 1.1 times
faster in any system than the combination of the pseudoparticle
multipole method and Phantom-GRAPE.  Our implementation allows
simulating homogeneous systems up to 2.2 times faster than that with
only monopole terms, however, 
speed up for clustered systems is not enough
because the increase of approximated interactions is insufficient to
negate the increased calculation cost by computing quadrupole terms.
We have estimated that improvement in performance can be achieved by
the use of a new SIMD instruction set, AVX-512.  Our code is expected
to be able to accelerate simulations of clustered systems up to 1.08
times faster on AVX-512 environment than that with only monopole
terms.

\end{abstract}

\section{Introduction}
Gravitational $N$-body simulations are widely used to study the
nonlinear evolution of astronomical objects such as 
the large-scale structure in the universe, 
galaxy clusters, galaxies, globular clusters, 
star clusters and planetary systems.

Directly solving $N$-body problems requires the computational cost in
proportion to $N^2$ and is unpractical for large $N$, where $N$ is the
number of particles.  Therefore, many ways to reduce the calculation
cost have been developed. 
One of the sophisticated algorithms is 
the tree method (\cite{key-1}) that evaluates gravitational forces with
calculation cost in proportion to $N\log{N}$.  The tree method
constructs a hierarchical oct-tree structure to represent a
distribution of particles and approximates the forces from a distant
group of particles by the multipole expansion. The opening parameter
$\theta$ is used to determine the tradeoff between accuracy and
performance.  If $l/d < \theta$, forces from a group of particles are
approximated by the multipole expansion, where $l$ is the spatial
extent of the group and $d$ is the distance to the group.  Thus,
larger $\theta$ gives higher performance and less accuracy.  

The tree
method is also used with the Particle-Mesh (PM) method (\cite{key-5})
when the periodic boundary condition is applied. This combination is
called as the TreePM method (\cite{key-13, key-14, key-19, key-20, key-12, key-6, key-21, key-25})
that calculates the short-range force by the tree method and the
long-range force by the PM method.  The TreePM method has been widely
used to follow the formation and evolution of the large-scale
structure in the universe and has been adopted in
many recent ultralarge cosmological $N$-body simulations.~(e.g., \cite{key-32})

For collisional $N$-body simulations that require high accuracy, the
Particle-Particle Particle-Tree (PPPT) algorithm (\cite{key-17}) has
been developed. In this algorithm, short-range forces are calculated
with the direct summation method and integrated with the fourth-order
Hermite method (\cite{key-29})
, and long-range forces are calculated with the tree
method and integrated with the leapfrog integrator.
The tree method has been combined with other algorithms and used to study
various astronomical objects.

Yet another way to accelerate $N$-body simulations is the use of
additional hardware, for example GRAPE (GRAvity PipE) systems
(\cite{key-22, key-7, key-8, key-33})
and
Graphics Processing Units (GPUs) (\cite{key-24, key-15, key-9, key-18, key-23}).
  GRAPEs are special-purpose hardware for gravitational
  $N$-body simulations and have been used to improve performances
  of $N$-body algorithms such as the tree (\cite{key-11}), and
  the TreePM (\cite{key-12}).

A different approach is utilizing a SIMD (Single Instruction Multiple
Data) instruction set. Phantom-GRAPE (\cite{key-10, key-2,
  key-3})~\footnote{https://bitbucket.org/kohji/phantom-grape} is a
highly-tuned software library and dramatically accelerates the
calculation of monopole terms utilizing a SIMD instruction set on x86
architecture. Quadrupole terms can be calculated by the combination of
the pseudoparticle multipole method (\cite{key-4}) and Phantom-GRAPE
for collisionless simulations (\cite{key-3}). 
In this method, a
quadrupole expansion is represented by three pseudoparticles.
However, the
pseudoparticle multipole method requires additional calculations such
as diagonalizations of quadrupole tensors that may cause substantial
performance loss.

To address this issue, we have implemented a software library that
accelerates the calculation of quadrupole terms by using a SIMD
instruction set AVX2 without positioning pseudoparticles. Our code is
based on Phantom-GRAPE for collisionless simulations and works as an
extension of the original Phantom-GRAPE. When the required accuracy is
the same, simulations should become faster by using quadrupole terms
than by using only monopole terms because we can increase the opening
angle $\theta$. Increasing $\theta$ gives another advantage that we
can also reduce the calculation cost of tree traversals.

The calculation including quadrupole terms should become further
efficient as the length of SIMD registers gets longer than 256-bit
(AVX2). Force evaluation is relatively scalable with respect to the
length. On the other hand, the time for tree traversals would not
be because hierarchical oct-tree structures are used.  Thus, in
environments such as AVX-512 with the SIMD registers of 512-bit
length, the total calculation for tree traversals and force evaluation
should be more accelerated in using quadrupole terms with larger
$\theta$ than in using the monopole only and smaller $\theta$.

This paper is organized as follows.  In section~\ref{sec:AVX}, we
overview the AVX2 instruction set.  We then describe the
implementation of our code in section~\ref{sec:imp}.  In
section~\ref{sec:acu} and \ref{sec:perf}, we show the accuracy and
performance, respectively.  Future improvement in performance by
utilizing AVX-512 is estimated in section~\ref{sec:dis}.
Section~\ref{sec:sum} is for the summary of this paper.

\section{The AVX2 instruction set}\label{sec:AVX}

The Advanced Vector eXtensions 2 (AVX2) is a SIMD
instruction set, which is an improved version of AVX.  Dedicated ``YMM
register'' with the 256-bit length is used to store eight single-precision
floating-point numbers or four double-precision floating-point
numbers.  The lower 128-bit of the YMM registers are called ``XMM
registers''.  The number of dedicated registers on a core is 16 in
AVX2.  Note that differently from AVX, AVX2 supports Fused
Multiply-Add (FMA) instructions for floating-point numbers.  More
precisely, AVX2 support and FMA support are not the same, but many
CPUs supporting AVX2 also support FMA instructions.

FMA instructions perform multiply-add operations.  Without FMA
instructions, a calculation $A \times B + C$ is done by two
operations, $D = A \times B$ and $D + C$.  With FMA instructions, this
calculation can be executed in one operation. Therefore, in such
situations, FMA instructions can gain the twice higher performance
than AVX environment.

Modern compilers do not necessarily generate optimized codes with
SIMD instructions from a source code written in high-level languages
because the detection of concurrency of loops and data dependency is
not perfect \citep{key-3}.  To manually assign YMM registers to
computational data in assembly-languages and use SIMD instructions
efficiently, we partially implemented our code with GCC (GNU Compiler
Collection) inline-assembly as original Phantom-GRAPE
\citep{key-10, key-2, key-3}.

\section{Implementation Details}\label{sec:imp}

In this section, we show our implementation that accelerates calculations
of quadrupole terms in the Barnes-Hut tree code utilizing the AVX2
instructions. 
Our code is
based on Phantom-GRAPE for collisionless simulations and works as an
extension of original Phantom-GRAPE (\cite{key-3}).
The quadrupole expansion of the potential at the
position $\mbox{\boldmath $r$}_i$ exerted by $n_j$ tree cells is expressed
as
\begin{eqnarray}\label{eq:pot}
\phi_i=&&-\sum_{j=1}^{n_j} \left\{\frac{Gm_j}{\sqrt{|\mbox{\boldmath $r$}_j - \mbox{\boldmath $r$}_i|^2 + \epsilon^2}}+ \right. \nonumber \\ 
&& \left. \frac{G}{2(|   \mbox{\boldmath $r$}_j - \mbox{\boldmath $r$}_i|^2 + \epsilon^2)^{5/2}}(\mbox{\boldmath $r$}_j - \mbox{\boldmath $r$}_i)\cdot\mbox{\boldmath $Q$}_j\cdot(\mbox{\boldmath $r$}_j - \mbox{\boldmath $r$}_i) \right\},
\end{eqnarray}
where $G, m_j, \mbox{\boldmath $r$}_j, \mbox{\boldmath $Q$}_j$, and
$\epsilon$ are the gravitational constant, the total mass of the
$j$-th cell, the position of the center of mass of the $j$-th cell, the
quadrupole tensor of the $j$-th cell, and the gravitational softening
length, respectively.
We represent the quadrupole tensor as
\begin{eqnarray}
\mbox{\boldmath $Q$}_j &=& \left[
\begin{array}{rrr}
q_{00} & q_{01} & q_{02} \\
q_{01} & q_{11} & q_{12} \\ 
q_{02} & q_{12} & q_{22} \\
\end{array}
\right] \nonumber \\
&=& \sum_{k=1}^{k_j}m_k\left[
\begin{array}{rrr}
3x_{jk}^2 - r_{jk}^2 & 3x_{jk}y_{jk} & 3x_{jk}z_{jk}\\
3y_{jk}x_{jk} & 3y_{jk}^2 - r_{jk}^2 & 3y_{jk}z_{jk}\\
3z_{jk}x_{jk} & 3z_{jk}y_{jk} & 3z_{jk}^2 - r_{jk}^2\\
\end{array}
\right], 
\end{eqnarray}
where $k_j$ is the number of particles in the $j$-th cell, $m_k$ is
the mass of the $k$-th particle, $x_k$, $y_k$ and $z_k$ are $x$, $y$
and $z$ component of the position of the $k$-th particle, $x_j$,
$y_j$, and $z_j$ are $x$, $y$, and $z$ component of the position of
the center of mass of the $j$-th cell, $x_{jk}=x_k-x_j$,
$y_{jk}=y_k-y_j$, $z_{jk}=z_k-z_j$, and $r_{jk}=\sqrt{x_{jk}^2 + y_{jk}^2
  + z_{jk}^2}$, respectively.  Since a quadrupole tensor is symmetric
and traceless, five values of $q_{00}, q_{01}, q_{02}, q_{11},$ and $q_{12}$ are
needed to memory a quadrupole tensor at least. The calculation of $q_{22}$ is as
\begin{equation}
q_{22}=-(q_{00}+q_{11}).
\end{equation}
However, our code loads the value of $q_{22}$ instead of calculating
to avoid redundant calculations of $q_{22}$ of the same cell.
Therefore, our code loads the six numbers to memory a quadrupole tensor.

The first term in the summation of the
equation~(\ref{eq:pot}) is the monopole term, and the second term is
the quadrupole. We rewrite the monopole term as $\phi_{j}^\mathrm{mono}$
and the quadrupole term as
$\phi_{j}^\mathrm{quad}$. 
These are 
\begin{equation}
\phi_{j}^\mathrm{mono} = \frac{Gm_j}{\hat{r}_{ij}},
\end{equation}
\begin{equation}
\phi_{j}^\mathrm{quad} = \frac{G}{2\hat{r}_{ij}^5}\mbox{\boldmath $r$}_{ij}\cdot\mbox{\boldmath $Q$}_j\cdot\mbox{\boldmath $r$}_{ij},
\end{equation}
where $\hat{r}_{ij}=\sqrt{|\mbox{\boldmath $r$}_j - \mbox{\boldmath $r$}_i|^2 + \epsilon^2}$, 
and $\mbox{\boldmath $r$}_{ij} = \mbox{\boldmath $r$}_j - \mbox{\boldmath $r$}_i$.
The gravitational force at the position $\mbox{\boldmath $r$}_i$ is given as follows:
\begin{equation}\label{eq:nabla}
\mbox{\boldmath $a$}_i = -\nabla\phi_i.
\end{equation}
From equation~(\ref{eq:pot}) and equation~(\ref{eq:nabla}),
\begin{equation}\label{eq:force}
\mbox{\boldmath $a$}_i = -\sum_{j=1}^{n_j}\left(\frac{\phi_{j}^\mathrm{mono} + 5\phi_{j}^\mathrm{quad}}{\hat{r}_{ij}^2}\mbox{\boldmath $r$}_{ij} - \frac{1}{\hat{r}_{ij}^5}\mbox{\boldmath $Q$}_j\cdot\mbox{\boldmath $r$}_{ij}\right).
\end{equation}

We aim to speed up the calculations of potential given in
equation~(\ref{eq:pot}) and a gravitational force given in
equation~(\ref{eq:force}) with AVX2 instructions. In those equations,
the $j$-th cell exerts forces on the $i$-th particle. In this paper,
we call them as ``$j$-cells'', and ``$i$-particles''.

Since forces exerted by $j$-cells on $i$-particles
are independent of each other, multiple forces
can be calculated in parallel. Since the AVX2 instructions compute eight single-precision floating-point numbers in parallel, 
our code calculates the forces on four $i$-particles from two $j$-cells in parallel 
as original Phantom-GRAPE~(\cite{key-3}).

\subsection{Structures for the particle and cell data}
The data assignment of four $i$-particles in YMM registers is the same
as original Phantom-GRAPE for collisionless simulations.  The data
assignment of two $j$-cells in YMM registers is also the same as the
assignment of two $j$-particles on original Phantom-GRAPE for
collisionless simulations.  The details are given in \citet{key-3}.

Our implementation shares the structures for $i$-particles, the
resulting forces, and potentials with original Phantom-GRAPE for
collisionless simulations. We define the structures for $j$-cells as
shown in List~1. The positions of the center of mass, total masses,
and quadrupole tensors of two $j$-cells are stored in the structure
\verb|Jcdata|.

\begin{lstlisting}[label=jcdata]
// List 1: Structure for j-cells
typedef struct jcdata{
  // xm={{x0, y0, z0, m0}, {x1, y1, z1, m1}}
  float xm[2][4];
  /*
    q={
    {q0-00, q0-01, q0-02, 0.0, 
     q1-00, q1-01, q1-02, 0.0},
    {q0-11, q0-12, q0-22, 0.0,
     q1-11, q1-12, q1-22, 0.0}
    }
  */
  float q[2][8];
} Jcdata, *cJcdata;
\end{lstlisting}

\subsection{Macros for inline assembly codes}
Original Phantom-GRAPE defines some preprocessor macros expanded into
inline assembly codes. We use these macros to write a force loop for
calculating gravitational force on four $i$-particles with evaluating
quadrupole expansions.
Descriptions of the macros used in our code are summarized in
Table~\ref{tab:macros}. The title of Table~\ref{tab:macros} and the
descriptions of the macros except for \verb|VPERM2F128|, 
\verb|VEXTRACTF128|, \verb|VSHUFPS|, \verb|VFMADDPS|, and \verb|VFNMADDPS|
 are adapted from \citet{key-3}.  
Operands
\verb|reg|, \verb|reg1|, \verb|reg2|, \verb|dest|, and \verb|dst| specify the data in XMM or
YMM registers, and \verb|mem| is data in the main memory or the cache
memory. The operand named \verb|imm| is an 8-bit number to control the behavior
of some operations.  More details of the AVX2
instructions are presented in Intel's 
website~\footnote{https://software.intel.com/en-us/isa-extensions}.

\begin{table*}
  \caption{Descriptions of the macros for inline assembly codes. One 'value' denotes a single-precision floating-point number.}
  \begin{tabularx}{\textwidth}{lX}
      \hline
      Macro & Description\\ 
      \hline
      \verb|VLOADPS(mem, reg)| & Load four or eight packed values in \verb|mem| to \verb|reg| \\
\verb|VSTORPS(reg, mem)| & Store four or eight packed values in \verb|reg| to \verb|mem| \\
\verb|VADDPS(reg1, reg2, dst)| & Add \verb|reg1| to \verb|reg2|, and store the result to \verb|dst| \\
\verb|VSUBPS(reg1, reg2, dst)| & Subtract \verb|reg1| from \verb|reg2|, and store the result to \verb|dst| \\
\verb|VMULPS(reg1, reg2, dst)| & Multiply \verb|reg1| by \verb|reg2|, and store the result to \verb|dst| \\
\verb|VRSQRTPS(reg, dst)| & Compute the inverse-square-root of \verb|reg|, and store the result to \verb|dst| \\
\verb|VZEROALL| & Zero all YMM registers \\
\verb|VPERM2F128(src1, src2, dest, imm)| & Permute 128-bit floating-point fields in \verb|src1| and \verb|src2| using controls from \verb|imm| 
, and store result in \verb|dest| \\
\verb|VEXTRACTF128(src, dest, imm)| & Extract 128 bits of packed values from \verb|src| and store results in \verb|dest| \\
\verb|VSHUFPS(src1, src2, dest, imm)| & Shuffle packed values selected by \verb|imm| from \verb|src1| and \verb|src2|, and store the result to \verb|dst| \\
\verb|PREFETCH(mem)| & Prefetch data on \verb|mem| to the cache memory \\
\verb|VFMADDPS(dst, reg1, reg2)| & Multiply eight packed values from \verb|reg1| and \verb|reg2|, add to \verb|dst| and put the result in \verb|dst|. \\
\verb|VFNMADDPS(dst, reg1, reg2)| & Multiply eight packed values from \verb|reg1| and \verb|reg2|, negate the multiplication result and add to \verb|dst| and put result in \verb|dst|. \\
      \hline
    \end{tabularx}\label{tab:macros}
\begin{tabnote}
The title of this table and the descriptions of the macros except for \verb|VPERM2F128|, \verb|VEXTRACTF128|, \verb|VSHUFPS|, \verb|VFMADDPS|, and \verb|VFNMADDPS| are adapted from \citet{key-3}.
\end{tabnote}
\end{table*}

\subsection{A force loop}
The following routine computes the forces on four $i$-particles from $j$-cells.
\begin{enumerate}
 \item Zero all the YMM registers.
 \item\label{step1} Load the $x$, $y$ and $z$ coordinates of four $i$-particles to the lower 128-bit of YMM00, YMM01 and YMM02, and copy them to the upper 128-bit of YMM00, YMM01 and YMM02, respectively.
 \item Load the $x$, $y$ and $z$ coordinates of the center of mass and the total masses of two $j$-cells to YMM14.
 \item Broadcast the $x$, $y$, and $z$ coordinates of the center of mass of two $j$-cells in YMM14 to YMM03, YMM04, and YMM05, respectively.
 \item Subtract YMM00, YMM01 and YMM02 from YMM03, YMM04, and YMM05, then store the results ($x_{ij}, y_{ij}$ and $z_{ij}$) in YMM03, YMM04, YMM05, respectively.
 \item\label{step6} Load squared softening lengths to the lower 128-bit of YMM01, and copy them to the upper 128-bit of YMM01.
 \item\label{step7} Square $x_{ij}$ in YMM03, $y_{ij}$ in YMM04, $z_{ij}$ in YMM05 and add them to the squared softening lengths in YMM01. 
It is the softened squared distances
$\hat{r}^2_{ij} \equiv r^2_{ij}+\epsilon^2$ between the center of mass
of two $j$-cells and four $i$-particles are stored in YMM01.
 \item\label{step8} Calculate inverse-square-root for $\hat{r}^2_{ij}$ in YMM01, and store the result $1/\hat{r}_{ij}$ in YMM01.
 \item\label{step9} Square $1/\hat{r}_{ij}$ in YMM01 and store the results in YMM00.
 \item\label{step10} Broadcast the total masses of two $j$-cells in YMM14 to YMM02.
 \item Multiply $1/\hat{r}_{ij}$ in YMM01 by $m_j$ in YMM02 to obtain $\phi_{j}^\mathrm{mono}=m_j/\hat{r}_{ij}$, and store the results in YMM02.
 \item Load $q_{00}$, $q_{01}$ and $q_{02}$ of two $j$-cells to YMM08, $q_{11}$, $q_{12}$ and $q_{22}$ of two j-cells to YMM15, respectively.
 \item Broadcast the $q_{00}$, $q_{01}$, $q_{02}$, $q_{11}$, $q_{12}$ and $q_{22}$ to YMM06, YMM07, YMM08, YMM13, YMM14, YMM15, respectively.
 \item Multiply YMM03, YMM04, and YMM05 by YMM06, YMM07, and YMM08, respectively, and sum them up. The results are $x$-component of $\mbox{\boldmath $Q$}_j\cdot\mbox{\boldmath $r$}_{ij}$, and stored in YMM06.
 \item Multiply YMM03, YMM04, and YMM05 by YMM07, YMM13, and YMM14, respectively, and sum them up. The results are $y$-component of $\mbox{\boldmath $Q$}_j\cdot\mbox{\boldmath $r$}_{ij}$, and stored in YMM13.
  \item Multiply YMM03, YMM04, and YMM05 by YMM08, YMM14, and YMM15, respectively, and sum them up. The results are $z$-component of $\mbox{\boldmath $Q$}_j\cdot\mbox{\boldmath $r$}_{ij}$, and stored in YMM15.
  \item Multiply YMM06, YMM13, and YMM15 by YMM03, YMM04, and YMM05, respectively, and sum them up to calculate $\mbox{\boldmath $r$}_{ij}\cdot\mbox{\boldmath $Q$}_j\cdot\mbox{\boldmath $r$}_{ij}$. The results are stored in YMM07.
 \item Square $1/\hat{r}^2_{ij}$ in YMM00 and store the results in YMM08.
 \item Multiply $1/\hat{r}^4_{ij}$ in YMM08 by $1/\hat{r}_{ij}$ in YMM01 to calculate $1/\hat{r}^5_{ij}$ and store the results in YMM08.
 \item Load 0.5 in YMM14.
 \item Multiply $\mbox{\boldmath $r$}_{ij}\cdot\mbox{\boldmath $Q$}_j\cdot\mbox{\boldmath $r$}_{ij}$ in YMM07 by $1/\hat{r}^5_{ij}$ in YMM08, then multiply it by 0.5 in YMM14 to calculate $\phi_{j}^\mathrm{quad}$ and store the results in YMM02.
 \item Accumulate $\phi_{j}^\mathrm{mono}$ in YMM02 and $\phi_{j}^\mathrm{quad}$ in YMM07 into $\phi_i$ in YMM09.
 \item Load 5 in YMM14.
 \item Calculate $\phi_{j}^\mathrm{mono} + 5.0\phi_{j}^\mathrm{quad}$ and store the results in YMM02.
 \item Multiply YMM00 by YMM03, YMM04, and YMM05 to calculate $x$, $y$, and $z$ components of the first term of the summation in equation~\ref{eq:force}, then accumulate them into YMM10, YMM11 and YMM12, respectively.
 \item Multiply YMM08 by YMM06, YMM13, and YMM15 to calculate $x$, $y$, and $z$ components of the second term of the summation in equation~\ref{eq:force}, then subtract them from YMM10, YMM11, and YMM12, respectively.
 \item Return to step \ref{step1} until all the j-cells are processed.
 \item Perform sum reduction of partial forces and potentials in the lower and upper 128-bits of YMM10, YMM11, YMM12, and YMM09, and store the results in the lower 128-bit of YMM10, YMM11, YMM12, YMM09, respectively.
 \item Store forces and potentials in the lower 128-bit of YMM10, YMM11, YMM12, and YMM09 to the structure \verb|Fodata|.
\end{enumerate}

List~2 is the function \verb|c_GravityKernel| calculating the forces on four
$i$-particles. We changed the order of operations in an actual code a
little to make contiguous instructions independently, resulting in
improved throughput.  The data of $i$-particles and the squared
softening length are common for all $j$-cells.  However, unlike
original Phantom-GRAPE for collisionless system, loading the data of
$i$-particles is necessary for each $j$ loop, because the number of
SIMD registers of AVX2 is not enough to keep the data over the loop.
In step~\ref{step6} squared softening lengths overwrite
$y$-coordinates of $i$-particles in YMM01 and are replaced with
$\hat{r}^2_{ij}$ in step~\ref{step7}.  In step~\ref{step9}
$x$-coordinates of $i$-particles in YMM00 are replaced with
$1/\hat{r}^2_{ij}$ .  In step~\ref{step10} $z$-coordinates of
$i$-particles in YMM02 are replaced with $m_j$.

Assuming that one division and one square-root each require 10
floating point operations~(\cite{key-26}), thus one
inverse-square-root requires 20 floating point operations.
The number of floating point operations
needed for the calculation of force exerted by one $j$-cell on one
$i$-particle is counted to be 71. According to IntelR 64 and IA-32
Architectures Optimization Reference
Manual~\footnote{https://www.intel.com/content/dam/doc/manual/64-ia-32-architectures-optimization-manual.pdf},
the latency of one inverse-square-root (\verb|VRSQRTPS|) is seven.
Therefore, if we assume that one inverse-square-root requires seven
floating point operations, the total number of floating point
operations per interaction is counted to be 58.

\begin{lstlisting}[label=list:force]
/* 
    List 2: A force loop which evaluates up 
    to quadrupole term by using AVX2.
*/
void c_GravityKernel(pIpdata ipdata,
                     pFodata fodata,
                     cJcdata jcdata, int nj){
  int j;
  float five[8] = {5.0, 5.0, 5.0, 5.0,
                   5.0, 5.0, 5.0, 5.0};
  float half[8] = {0.5, 0.5, 0.5, 0.5,
                   0.5, 0.5, 0.5, 0.5};
  PREFETCH(jcdata[0]);
  
  VZEROALL;
  for(j = 0; j < nj; j += 2){
    // load i-particle
    VLOADPS(*ipdata->x, XMM00);
    VLOADPS(*ipdata->y, XMM01);
    VLOADPS(*ipdata->z, XMM02);
    VPERM2F128(YMM00, YMM00, YMM00, 0x00);
    VPERM2F128(YMM01, YMM01, YMM01, 0x00);
    VPERM2F128(YMM02, YMM02, YMM02, 0x00);
    // load jcell's coordinate
    VLOADPS(jcdata->xm[0][0], YMM14);
    VSHUFPS(YMM14, YMM14, YMM03, 0x00); //xj
    VSHUFPS(YMM14, YMM14, YMM04, 0x55); //yj
    VSHUFPS(YMM14, YMM14, YMM05, 0xaa); //zj
    // r_ij,x -> YMM03
    VSUBPS(YMM00, YMM03, YMM03);
    // r_ij,y -> YMM04
    VSUBPS(YMM01, YMM04, YMM04);
    // r_ij,z -> YMM05
    VSUBPS(YMM02, YMM05, YMM05);
    // eps^2 -> YMM01
    VLOADPS(*ipdata->eps2, XMM01);
    VPERM2F128(YMM01, YMM01, YMM01, 0x00);
    // r_ij^2 -> YMM01
    VFMADDPS(YMM01, YMM03, YMM03); 
    VFMADDPS(YMM01, YMM04, YMM04);
    VFMADDPS(YMM01, YMM05, YMM05);
    // 1 / r_ij -> YMM01
    VRSQRTPS(YMM01, YMM01);
    // 1 / r_ij^2 -> YMM00
    VMULPS(YMM01, YMM01, YMM00);
    // phi_p(mj / r_ij) -> YMM02
    VSHUFPS(YMM14, YMM14, YMM02, 0xff); // mj
    VMULPS(YMM01, YMM02, YMM02);

    /*
       q00, q01, q02, q11, q12, q22
       -> YMM06, 07, 08, 13, 14, 15, 
       respectively
    */
    VLOADPS(jcdata->q[0][0], YMM08);
    VLOADPS(jcdata->q[1][0], YMM15);
    VSHUFPS(YMM08, YMM08, YMM06, 0x00);
    VSHUFPS(YMM08, YMM08, YMM07, 0x55);
    VSHUFPS(YMM08, YMM08, YMM08, 0xaa);
    VSHUFPS(YMM15, YMM15, YMM13, 0x00);
    VSHUFPS(YMM15, YMM15, YMM14, 0x55);
    VSHUFPS(YMM15, YMM15, YMM15, 0xaa);

    // q00 * r_ij,x -> YMM06
    VMULPS(YMM03, YMM06, YMM06);
    // YMM06 + q01 * r_ij,y -> YMM06
    VFMADDPS(YMM06, YMM04, YMM07);
    // YMM06 + q02 * r_ij,z -> YMM06
    VFMADDPS(YMM06, YMM05, YMM08);

    // q11 * r_ij,y -> YMM13
    VMULPS(YMM13, YMM04, YMM13);
    // YMM13 + q01 * r_ij,x -> YMM13
    VFMADDPS(YMM13, YMM03, YMM07);
    // YMM13 + q12 * r_ij,z -> YMM13
    VFMADDPS(YMM13, YMM05, YMM14);

    // q22 * r_ij,z -> YMM15
    VMULPS(YMM15, YMM05, YMM15);
    // YMM15 + q02 * r_ij,x -> YMM15
    VFMADDPS(YMM15, YMM03, YMM08);
    // YMM15 + q12 * r_ij,y -> YMM15
    VFMADDPS(YMM15, YMM04, YMM14); 
    
    // calculate drqdr
    // qdr[0] * r_ij,x -> YMM07
    VMULPS(YMM03, YMM06, YMM07);
    // YMM07 + qdr[1] * r_ij,y -> YMM07
    VFMADDPS(YMM07, YMM04, YMM13);
    // YMM07 + qdr[2] * r_ij,z -> YMM07
    VFMADDPS(YMM07, YMM05, YMM15);

    // 1/(r_ij)^5 -> YMM08
    VMULPS(YMM00, YMM00, YMM08);
    VMULPS(YMM01, YMM08, YMM08); 

    // 0.5 -> YMM14
    VLOADPS(half, YMM14);
    // 1/(r_ij)^5 * drqdr * 0.5 -> YMM07
    VMULPS(YMM07, YMM08, YMM07);
    VMULPS(YMM07, YMM14, YMM07);

    // phi += phi_p(YMM02) + phi_q(YMM07)
    VADDPS(YMM02, YMM07, YMM14);
    VADDPS(YMM14, YMM09, YMM09);
    // 5.0 -> YMM14
    VLOADPS(five, YMM14);
    // 5.0 * phi_q + phi_p -> YMM02
    VFMADDPS(YMM02, YMM07, YMM14);

    // YMM02 / (r_ij)^2 ->YMM00
    VMULPS(YMM02, YMM00, YMM00); 

    // ax, ay, az -> YMM10, YMM11, YMM12
    VFMADDPS(YMM10, YMM00, YMM03);
    VFMADDPS(YMM11, YMM00, YMM04);
    VFMADDPS(YMM12, YMM00, YMM05);
    VFNMADDPS(YMM10, YMM08, YMM06);
    VFNMADDPS(YMM11, YMM08, YMM13);
    VFNMADDPS(YMM12, YMM08, YMM15);

    jcdata++;
  }
  VEXTRACTF128(YMM10, XMM00, 0x01);
  VEXTRACTF128(YMM11, XMM01, 0x01);
  VEXTRACTF128(YMM12, XMM02, 0x01);
  VEXTRACTF128(YMM09, XMM03, 0x01);
  VADDPS(YMM10, YMM00, YMM10);
  VADDPS(YMM11, YMM01, YMM11);
  VADDPS(YMM12, YMM02, YMM12);
  VADDPS(YMM09, YMM03, YMM09);

  VSTORPS(XMM10,  *fodata->ax);
  VSTORPS(XMM11,  *fodata->ay);
  VSTORPS(XMM12,  *fodata->az);
  VSTORPS(XMM09, *fodata->phi);
}
\end{lstlisting}

\subsection{Application programming interfaces}
List~3 shows the application programming interfaces~(APIs) for our
code.
\verb|g5c_set_nMC| tells our code the number of $j$-cells.
\verb|g5c_set_xmjMC| transfer positions, mass and quadrupole tensors
of $j$-cells to the array of the structure \verb|Jcdata|.
\verb|g5c_calculate_force_on_xMC| transmits coordinates and number of
$i$-particles to an array of the structure \verb|Ipdata|, which is
defined in the original Phantom-GRAPE~(\cite{key-3}), and calculates
the forces and potentials exerted by $j$-cells on the $i$-particles
and store the result in the arrays \verb|ai| and \verb|pi|,
respectively.

List~4 shows a part of C++ code that calculates the forces and potentials
of all particles. In this code, we use the modified tree
algorithm~(\cite{key-27}), where 
the particles in a cell that contains
$n_{\rm crit}$ or less particles shares the same interaction list.
The particles
sharing the same interaction list are $i$-particles, the particles
in the interaction list are $j$-particles, and the cells in the
interaction list are $j$-cells. The functions beginning with
\verb|g5_| are the APIs for the original Phantom-GRAPE~(\cite{key-3}),
and calculate particle-particle interactions. The functions beginning
with \verb|g5c_| are the APIs for our code, and calculate interactions
from cells.

\begin{lstlisting}
// List 3: APIs for our code.
void g5c_set_xmjMC(int devid, int adr,
            int nj, double (*xj)[3],
            double *mj, double (*qj)[6]);
void g5c_set_nMC(int devid, int n);
void g5c_calculate_force_on_xMC(int devid, 
            double (*x)[3], double (*a)[3],
            double *p, int ni);
\end{lstlisting}

\begin{lstlisting}[label=sample]
// List 4: Sample code
class particle; // Contains particle data

class node; // Contains cell data

/*
  a cell that contain particles which 
  share same interactions
*/
class ilist{
public:
  int ni; // Number of particles
  double l; // Cell's length
  double (*xi)[3]; // Position
  double (*ai)[3]; // Force
  double (*pi); // Potential
  particle*(*pp); // Pointer to particle
  double cpos[3]; // Cell's center
};

class jlist{// contains one j-particle data
public:
  int nj; // Number of particles
  double (*xj)[3]; // Position
  double (*mj); // Mass
};

class jcell{// contains one j-cell data
public:
  int nj; // Number of cells
  double (*xj)[3]; // Mass center
  double (*mj); // Total mass
  double (*qj)[6]; // Quadrupole tensor
};

/*
  create tree structure and groups of 
  i-particles which share the same 
  interaction list.
*/
void create_tree(node *, particle *, int,
                 ilist, int, int);

/*
  traverse the tree structure 
  and make lists of j-particles and j-cells.
  (a interaction list.)
*/
void traverse_tree(node *, ilist, jlist,
                   jcell, double, int);

/*
  assign or add the values of force and 
  potential in ilist to those in
  particle class.
*/
void assign_force_potential(ilist);
void add_force_potential(ilist);

int n; // number of particles
double theta2; // square of theta

/* 
   calculate forces and potentials of 
   all particles.
*/
void calc_force(int n, int nnodes, 
                particle pp[], node *bn,
                double eps, double theta2,
                int ncrit){
  // Number of groups of i-particles
  int ni;
  // index of loop
  int i, k;

  create_tree(bn, pp, ni, i_list, n, ncrit);

  g5_open();
  g5_set_eps_to_all(eps);
  
  for(i = 0; i < ni; i++){
    tree_traversal(bn, i_list, j_list, 
                   j_cell, theta2, ncrit);

    /*
      calculate forces exerted by 
      j-particles
    */
    g5_set_xmjMC(0, 0, j_list->nj,
                 j_list->xj, j_list->mj);
    g5_set_nMC(0, j_list->nj);
    g5_calculate_force_on_xMC(0, 
                              i_list[i]->xi,
                              i_list[i]->ai,
                              i_list[i]->pi,
                              i_list[i]->ni
                             );
    assign_force_potential(i_list);

    // calculate forces exerted by j-cells
    g5c_set_xmjMC(0, 0, j_cell->nj,
                  j_cell->xj, j_cell->mj,
                  j_cell->qj);
    g5c_set_nMC(0, j_cell->nj);
    g5c_calculate_force_on_xMC(0,
                               i_list[i]->xi,
                               i_list[i]->ai,
                               i_list[i]->pi,
                               i_list[i]->ni
                              );
    add_force_potential(i_list);
  }
  g5_close();
}
\end{lstlisting}

\section{Accuracy}\label{sec:acu}
\begin{figure}
\begin{center}
\includegraphics[width=8cm]{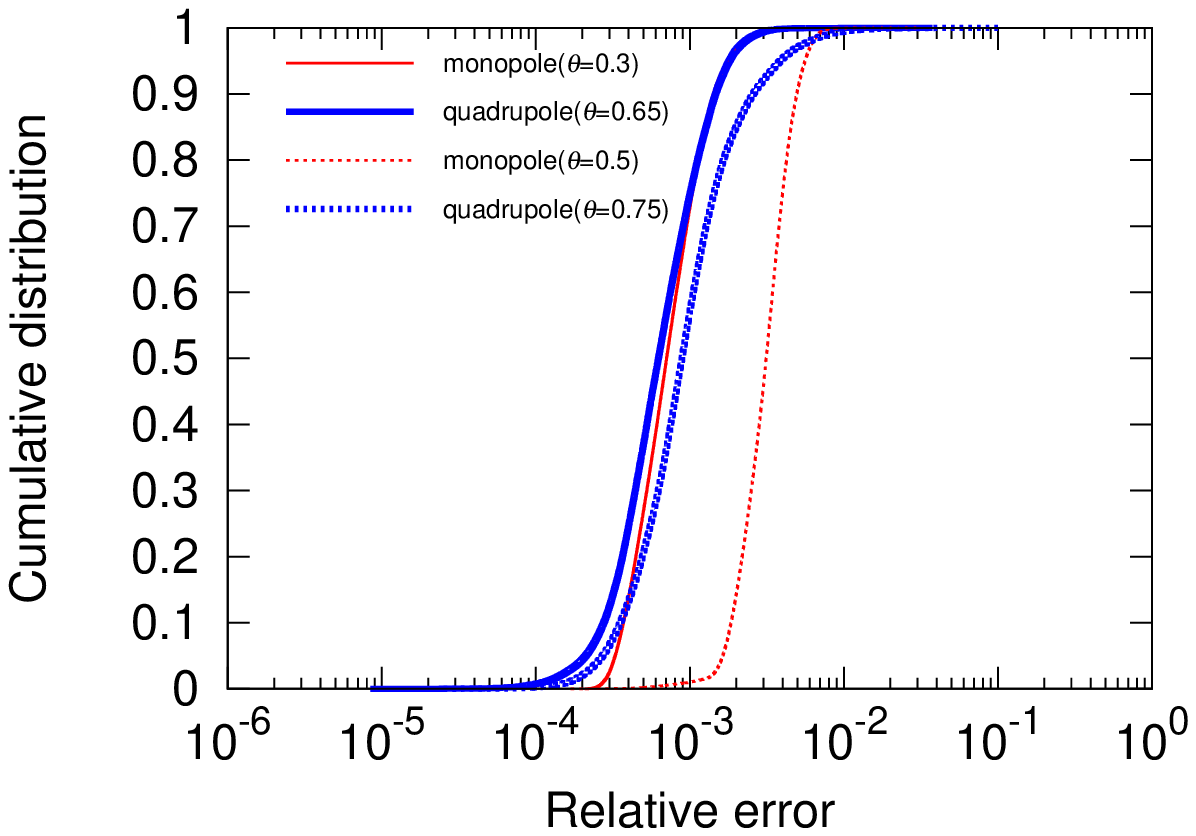}
\includegraphics[width=8cm]{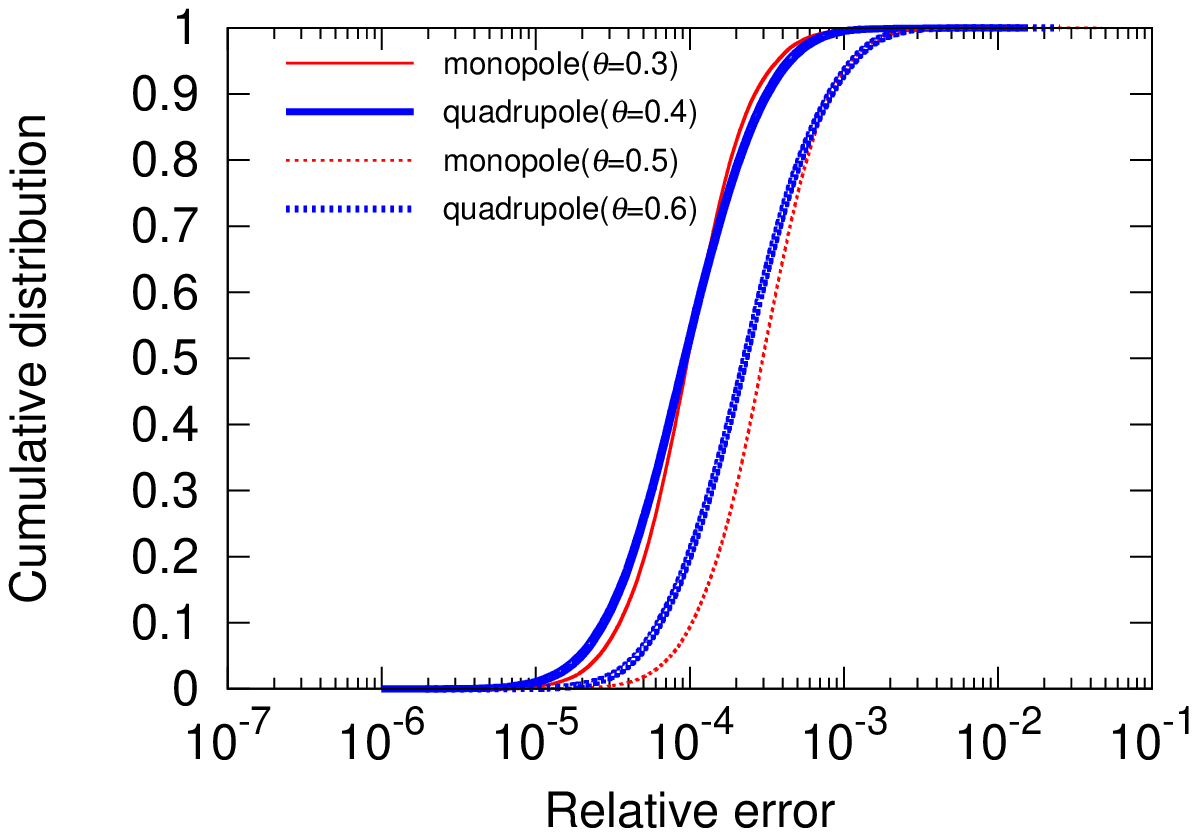}
\includegraphics[width=8cm]{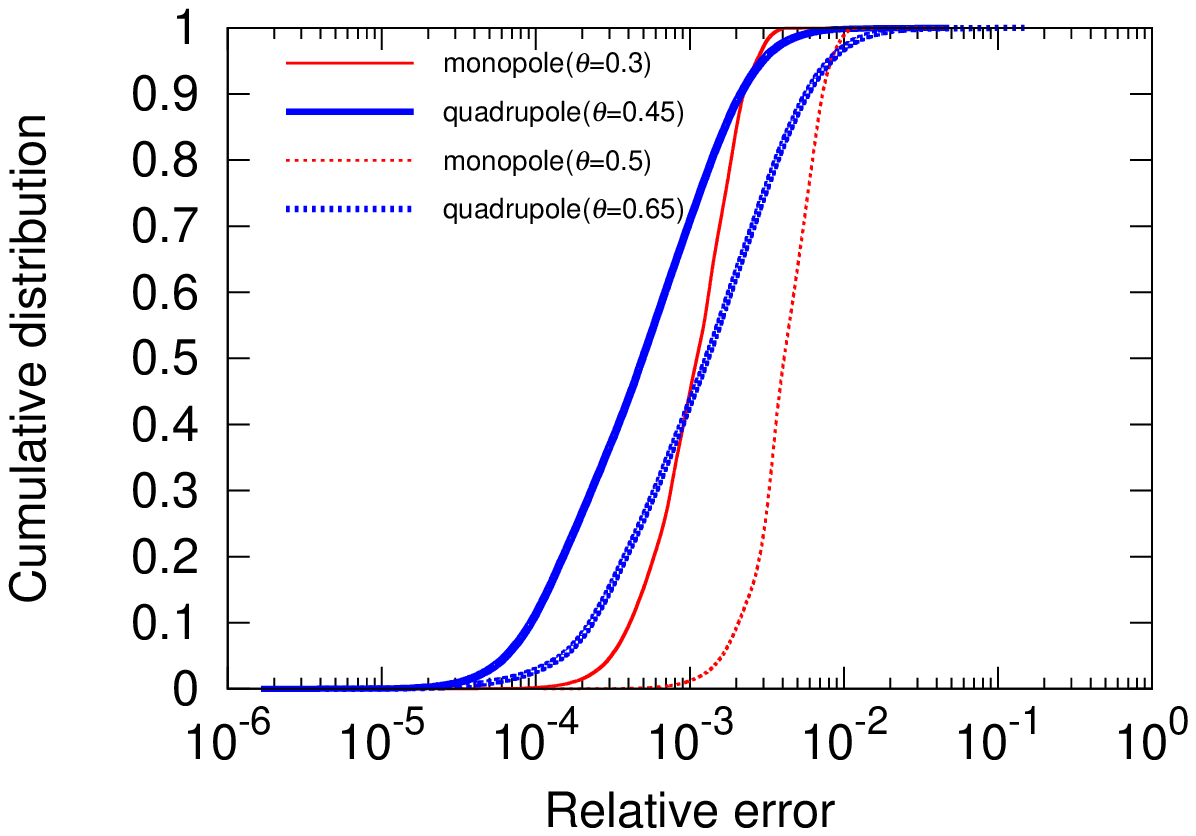}
\caption{Cumulative distribution of errors in forces of particles with
  $N=65,536$. From top to bottom, 
the particle distributions are 
a homogeneous sphere, a Plummer model and an exponential disk, 
respectively.
}
\label{fig:err_cc65k}
\end{center}
\end{figure}

In this section, we compare the accuracy of forces obtained by utilizing only monopole terms and that obtained by calculating up to quadrupole terms. The detailed discussion about errors of forces in the tree method is given in \citet{key-30}, Barnes and Hut~(\yearcite{key-31}), and \citet{key-28}. 
Figure~\ref{fig:err_cc65k} shows the cumulative distribution of
relative force errors in particles distributed in a
homogeneous sphere (top), a Plummer model (middle), and an exponential
disk (bottom), respectively. Relative errors in the forces of
particles are given as
\begin{eqnarray}
\frac{|\mbox{\boldmath $a$}_{\mathrm{TREE}}-\mbox{\boldmath $a$} _{\mathrm{DIRECT}}|}{|\mbox{\boldmath $a$}_{\mathrm{DIRECT}}|} \label{eq:error_acc} ,
\end{eqnarray}
where $\mbox{\boldmath $a$}_{\mathrm{TREE}}$ is the force calculated
using the tree method, and $\mbox{\boldmath $a$}_{\mathrm{DIRECT}}$ is
the force computed using the direct particle-particle method with
Phantom-GRAPE for collisionless simulations.  We used our
implementation to calculate quadrupole terms and original
Phantom-GRAPE to calculate monopole terms. The number of particles
is 65,536 for all three particle distributions.

The top panel of Figure~\ref{fig:err_cc65k} (the homogeneous sphere)
shows that the result of using quadrupole terms with $\theta=0.65$ has
accuracy comparable to that of only monopole terms with $\theta=0.3$.
When using quadrupole terms with $\theta=0.75$, most particles have
smaller 
errors than using only monopole terms with $\theta=0.5$ and only
a few percent of particles have larger errors.

The middle 
panel (the Plummer model) of Figure~\ref{fig:err_cc65k}
suggests that about a half of the particles have 
smaller
errors with
calculating the quadrupole terms using $\theta=0.4$ than with
calculating only monopole terms using $\theta=0.3$. The rest of the
particles have slightly larger errors. However, these differences are
small and both error distributions agree with each other. The
result of using quadrupole terms with $\theta=0.6$ has accuracy
comparable to that of only monopole terms with $\theta=0.5$.  About a
tenth part of particles have larger errors when we calculate the
quadrupole terms with $\theta=0.6$ than when we calculate only the
monopole terms with $\theta=0.5$.

The bottom panel (the exponential disk) of Figure~\ref{fig:err_cc65k}
shows that the result of using quadrupole terms with $\theta=0.45$ has
accuracy comparable to that of only monopole terms with $\theta=0.3$.
When using quadrupole terms with $\theta=0.65$, most particles have
smaller 
errors than using only monopole terms with $\theta=0.5$ and only
a few percent of particles have larger errors.

In a homogeneous system, the net force exerted by particles located at
a certain range $r$ does not depend on $r$ because the gravitational
force from a particle at $r$ is proportional to $r^{-2}$ and the
number of particles at $r$ is proportional to $r^{2}$.  The force from
distant particles, which is not negligible compared to the force from
close particles, can be significantly more accurate by using
quadrupole than by using only the monopole.  On the other hand, in a
clustered system such as a Plummer model and a disk, the gravitational
force is dominated by nearby particles for a large fraction of
particles.  Thus accuracy cannot be significantly improved even if
quadrupole terms are used. Therefore, $\theta$ cannot be very large in
a clustered system.

\begin{figure}
\begin{center}
\includegraphics[width=8cm]{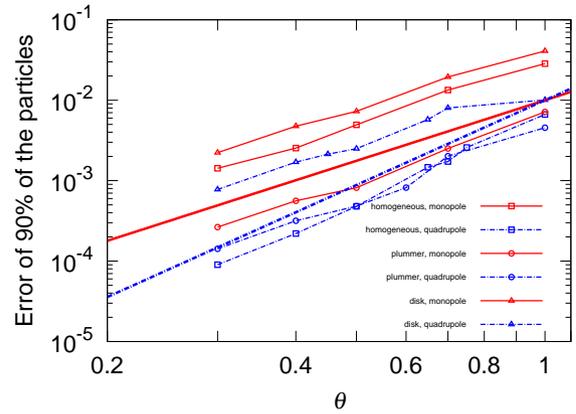}
\caption{ Error of 90\% of the particles as a function of
  $\theta$. The squares, circles, and triangles show the result of a
  homogeneous sphere, a Plummer model, and a disk, respectively. The
  solid and dashed 
lines without points show $\theta^{5/2}$ and
  $\theta^{7/2}$ scaling, respectively.
}
\label{fig:err_f90}
\end{center}
\end{figure}

Figure~\ref{fig:err_f90} shows the error at 90\% of the particles as a
function of $\theta$ and highlights the results described above. 
The errors when we utilize up to monopole terms and quadrupole terms are
roughly proportional to $\theta^{5/2}$ and $\theta^{7/2}$,
respectively.  This result is consistent with the scaling law of error
described in \citet{key-28}.
When we use only monopole terms, the error is the smallest
in the Plummer model because the net force is dominated by the forces
from nearby particles, most of which are calculated directly. The error
in the disk is the largest because of the anisotropic structure of the
disk.  If the same $\theta$ is used, calculation of quadrupole terms
reduces the error more in the homogeneous sphere than in other models
because the force from distant particles can be well approximated by
the quadrupole terms and such force constitutes a larger portion of the net
force in a homogeneous system than in a clustered system such as the
Plummer model and the disk.

\section{Performance}\label{sec:perf}
In this section, we compare the performance of our implementation, 
original Phantom-GRAPE, and the pseudoparticle multipole method 
when the same force accuracy is imposed. 
The system we used to measure the performance is shown in Table
\ref{tab:system}. We used only one core, and Intel Turbo Boost
Technology is enabled. Compiler options were -O3 -ffast-math
-funroll-loops. Theoretical peak FLOPS of the system per core is
67.2~GFLOPS. The values of $\theta$ when we utilize quadrupole moments
are based on the result that we described in
section~\ref{sec:acu}.

\begin{table}
\caption{The system we use to measure the performance.}
\begin{tabular}{cc}
  \hline
      CPU & Intel Xeon E5-2683 v4 2.10GHz \\ 
      Memory & 128GB \\
      OS &  CentOS Linux release 7.3.1611 (core) \\
      Compiler & gcc 4.8.5 20150623 (Red Hat 4.8.5-11) \\
      \hline
      \end{tabular}\label{tab:system}
\end{table}

\subsection{Comparison of calculation time when the same accuracy is required}
Table \ref{tab:LTsample} shows the wall clock time for evaluating
forces and potentials of all the particles with $N=4,194,304$. In
general, when we utilize quadrupole moments, the time consumed in the
tree construction becomes slightly longer because quadrupole tensors of
cells are calculated. When we use the pseudoparticle multipole method,
the time consumed in the tree construction becomes longer because of
the positioning of pseudoparticles.

The simulations of the homogeneous sphere with only the monopole
moments can be accelerated from 1.23 to 2.20 times faster when we use
our code and evaluate quadrupole terms. The simulations of the
exponential disk using only the monopole terms with $\theta=0.3$ can
be accelerated 1.13 times faster when we use our implementation and
set $\theta=0.45$. In other $\theta$ and particle distribution, using
the quadrupole terms slows simulations. As described in
section~\ref{sec:acu}, using quadrupole terms allows us to use
significantly larger $\theta$ than using only the monopole in a
homogeneous system, while we can increase $\theta$ moderately in a
clustered system. Therefore, more interactions from particles are
approximated by quadrupole expansion in a homogeneous system than in a
clustered system. Thus, using the quadrupole terms can efficiently
accelerate simulations of a homogeneous system. In the clustered system
such as the disk and the Plummer model, the number of approximated
interactions by using quadrupole terms and larger $\theta$ is not
enough to negate the increased calculation cost by computing
quadrupole terms.

Our implementation is always faster than the combination of pseudoparticle
multipole method and Phantom-GRAPE for collisionless simulations by a factor of 1.1 in any condition
because calculations such as diagonalizations of quadrupole tensors 
are unnecessary.

\begin{longtable}{cccccccc}
  \caption{
Wall clock time for evaluating forces and potentials of all the
particles with $N=4,194,304$. “Monopole” calculates only monopole
terms. “Pseudoparticle” calculates quadrupole terms with
pseudoparticles. “Quadrupole” calculates quadrupole terms with our
implementation. ``Homogeneous'' is the homogeneous sphere. ``Plummer''
is the Plummer model. ``Disk'' is the exponential
disk. $T_\mathrm{construct}$, $T_\mathrm{traverse}$, and
$T_\mathrm{force}$ are time for tree constructions, tree traverse,
force calculation, respectively. $T_\mathrm{total}$ is total time. The
column "Ratio" is ratios of the total time to that of using only
monopole.}\label{tab:LTsample} \hline Program & $\theta$ & Particle
  distribution & $T_\mathrm{construct}$[s] & $T_\mathrm{traverse}$[s]
  & $T_\mathrm{force}$[s] & $T_\mathrm{total}$[s] &
  Ratio\\ \endfirsthead \hline Program & $\theta$ & Particle
  distribution & $T_\mathrm{construct}$[s] & $T_\mathrm{traverse}$[s]
  & $T_\mathrm{force}$[s] & $T_\mathrm{total}$[s] & Ratio\\ \endhead
  \hline \endfoot \hline \endlastfoot \hline monopole & 0.3
  &Homogeneous& 1.06 & 3.03 & 19.82 & 23.99 & 1\\ pseudoparticle &
  0.65 &Homogeneous& 1.67 & 1.20 & 8.96 & 11.91 & 0.50 \\ quadrupole &
  0.65 &Homogeneous& 1.10 & 1.16 & 8.53 & 10.88 & 0.45\\ \hline
  monopole & 0.5 &Homogeneous& 1.06 & 1.34 & 7.77 & 10.26 &
  1\\ pseudoparticle & 0.75 &Homogeneous& 1.67 & 0.94 & 6.53 & 9.22 &
  0.90\\ quadrupole & 0.75 &Homogeneous& 1.10 & 0.92 & 6.25 & 8.35 &
  0.81\\ \hline monopole & 0.3 & Plummer & 2.05 & 7.79 & 31.30 & 41.23
  & 1\\ pseudoparticle & 0.4 & Plummer & 2.70 & 6.41 & 39.41 & 48.61 &
  1.18\\ quadrupole & 0.4 & Plummer & 2.11 & 5.81 & 36.64 & 44.65 &
  1.08\\ \hline monopole & 0.5 & Plummer & 2.05 & 2.71 & 10.28 & 15.13
  & 1\\ pseudoparticle & 0.6 & Plummer & 2.69 & 3.22 & 18.31 & 24.30 &
  1.61\\ quadrupole & 0.6 & Plummer & 2.11 & 2.96 & 17.11 & 22.26 &
  1.47\\ \hline monopole & 0.3 &Disk & 1.45 & 5.49 & 20.95 & 28.01 &
  1\\ pseudoparticle & 0.45 &Disk & 2.10 & 3.76 & 21.11 & 27.09 &
  0.97\\ quadrupole & 0.45 &Disk & 1.51 & 3.52 & 19.72 & 24.86 &
  0.89\\ \hline monopole & 0.5 &Disk & 1.45 & 2.22 & 7.96 & 11.75 &
  1\\ pseudoparticle & 0.65 &Disk & 2.11 & 1.98 & 9.70 & 13.92 &
  1.18\\ quadrupole & 0.65 &Disk & 1.51 & 1.88 & 9.18 & 12.67 & 1.08\\
\end{longtable}

\subsection{The dependency of calculation time in the number of particles and interactions per second}
\begin{figure}
 \begin{center}
\includegraphics[width=8cm]{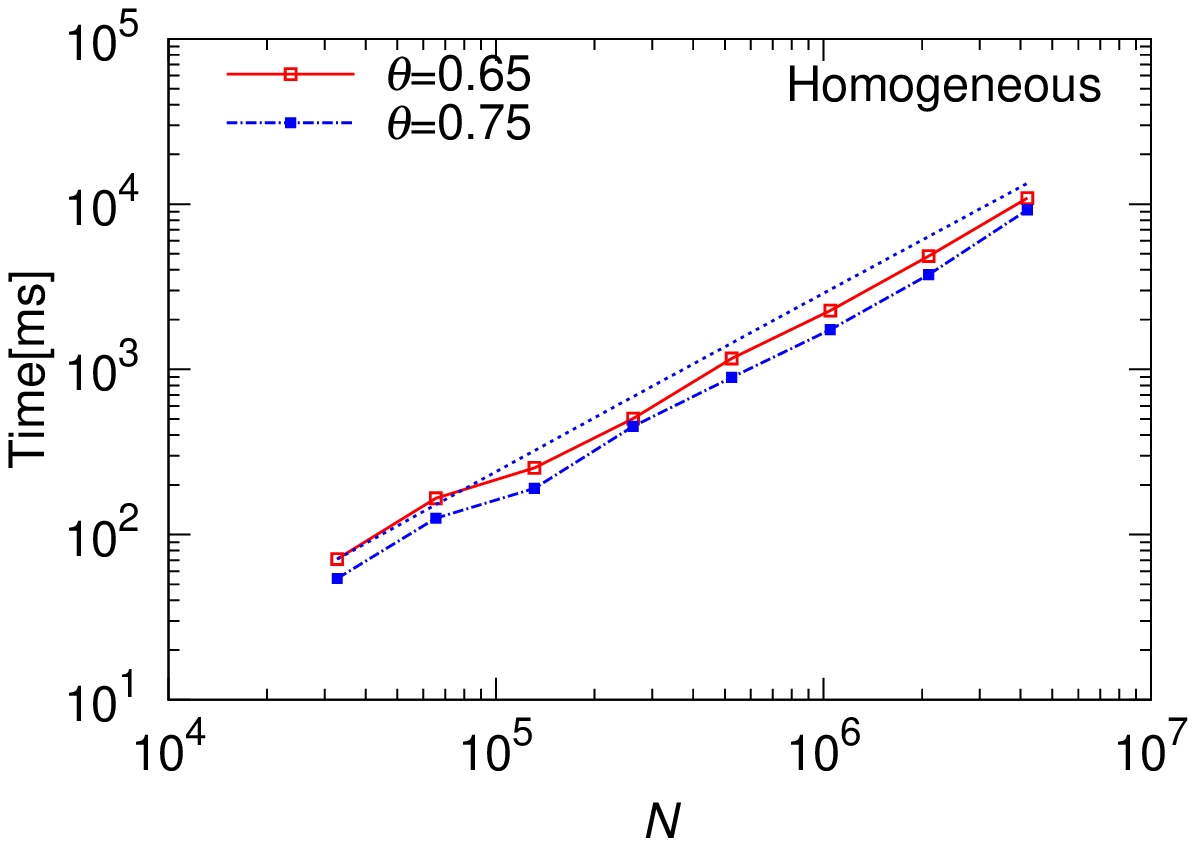}
\includegraphics[width=8cm]{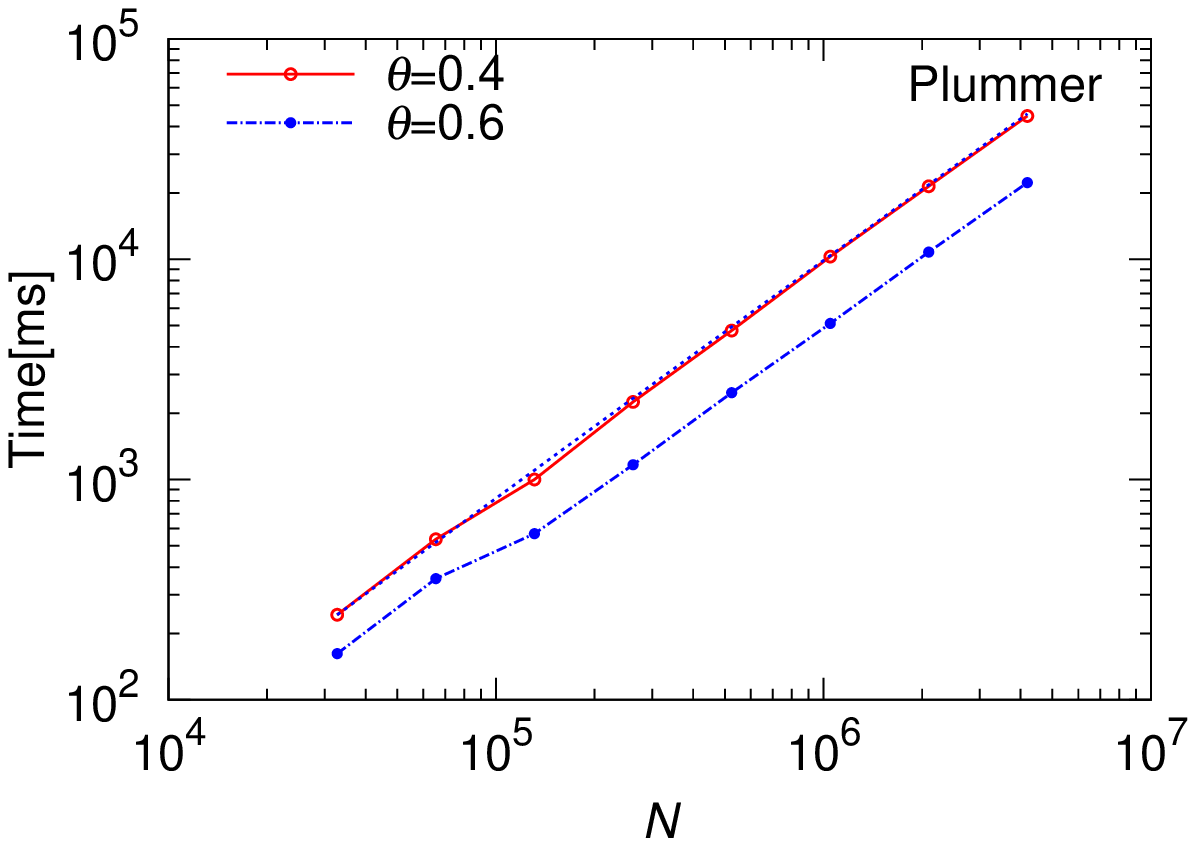}
\includegraphics[width=8cm]{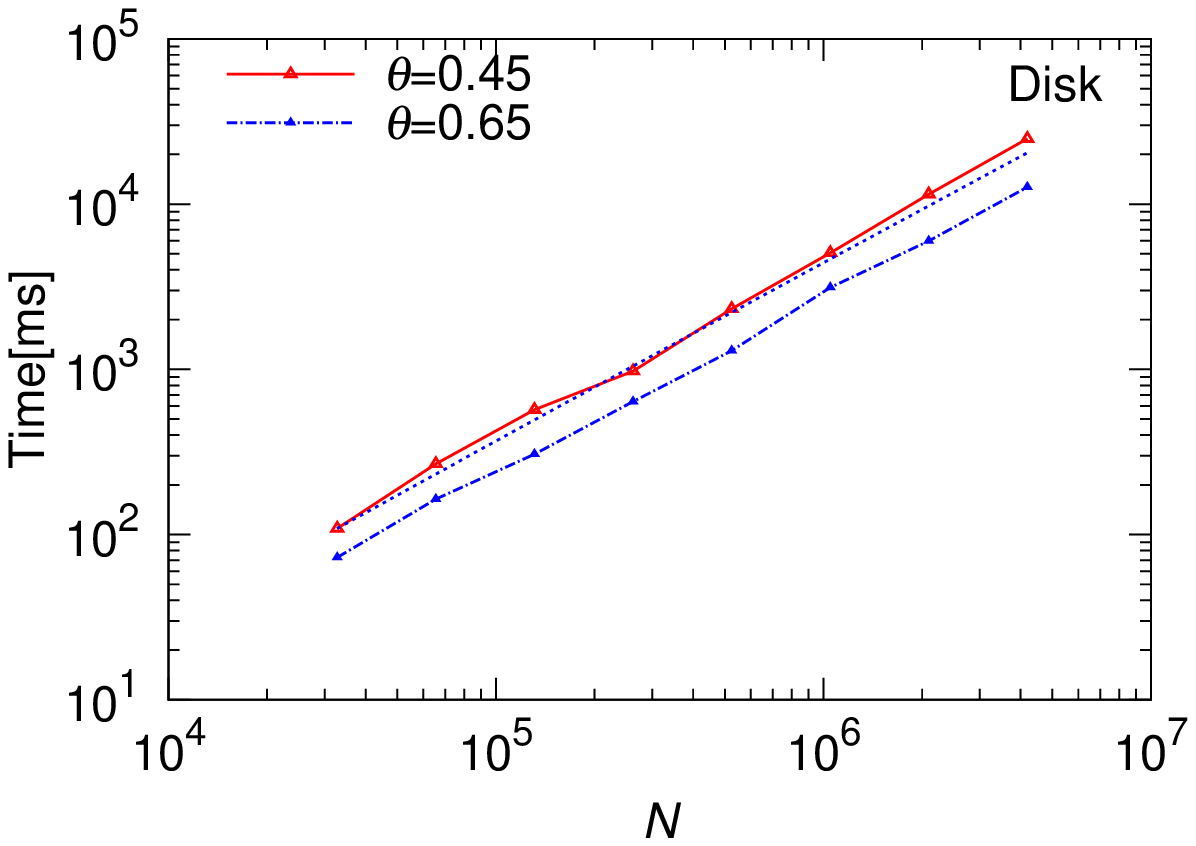} 
 \end{center}
\caption{Wall clock time for calculating forces and potentials of all
  the particles. From top to bottom, 
the particle
  distributions are a homogeneous sphere, a Plummer model and an
  exponential disk, respectively. Dotted lines show $N\log N$
  scaling.}\label{fig:n_cc}
\end{figure}

Figure \ref{fig:n_cc}
shows wall clock time on various $N$ for calculating forces and
potentials of particles in the homogeneous sphere 
(top),
the Plummer model 
(middle), 
and the exponential disk (bottom), respectively. Solid
curves are for small $\theta$, and dashed curves with points are for
large $\theta$. Dashed lines without point show $N \log N$ scaling. We
can see that the total time to calculate the force and potential of
particles is roughly proportional to $N \log N$. 
However, from
$N=65,536$ to $N=131,072$ on the homogeneous sphere, the actual scaling of the total time slightly deviates from the $N\log N$ scaling.
From $N=65,536$ to $N=131,072$, the depth level of the tree traversals
became deep because of the nature of the hierarchical oct-tree structure.
Thus, more part of interactions is approximated with the
multipole expansions. Therefore, the total number of particle-particle
and particle-cells interactions and the total calculation time
deviates slightly from the $N\log N$ scaling.
Deviation from $N \log N$ scaling can also be seen on the Plummer model and the disk. However, the deviation is not as obvious as that on the homogeneous sphere. The calculation time can fluctuate by other running processes.

\begin{figure}
 \begin{center}
  \includegraphics[width=8cm]{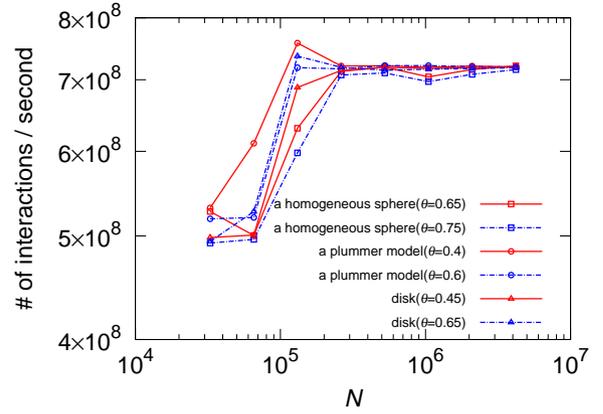} 
 \end{center}
\caption{
Numbers of interactions from cells per second. Note that
particle-particle interactions are excluded. The squares, circles, and
triangles show the result of a homogeneous sphere, a Plummer model,
and a disk, respectively. Solid curves show the result for smaller
$\theta$. Dotted curves show the result for larger $\theta$.}
\label{fig:ips}
\end{figure}

As seen in Figure~\ref{fig:ips}, the number of interactions from cells
per second is greatly reduced in $N < 262,144$.  This slowdown comes
from the overhead of storing $i$-particles into the structure named
\verb|Ipdata|.  Our code, as well as the original
Phantom-GRAPE~(\cite{key-3}) stores four $i$-particles into
\verb|Ipdata|. Each time a calculation of the net force on four
$i$-particles is done, next four $i$-particles are loaded into
\verb|Ipdata|.  The number of interaction is proportional to $n_i
\times n_j$, where $n_i$ is the number of $i$-particles, and the
computational cost for storing $i$-particles is proportional to
$n_i$. If $N$ becomes fewer, $n_i$ and $n_j$ also become
fewer. Therefore, the overhead of storing $i$-particles becomes
relatively large compared to the calculation of interactions itself,
resulting in the speed down of the calculation of interactions. This
behavior is also seen in the original Phantom-GRAPE~(\cite{key-3}),
which shows lower performance for smaller $n_i$ and $n_j$.

Theoretical peak FLOPS per core of the CPU which we use is 67.2 GFLOPS,
however, this value is based on the assumption that the CPU is
executing FMA operations all the time.  Actually, 36 counts of floating
point operations in our code are FMA , and the rest come from non-FMA,
add, subtract, multiply, and inverse-square root
operations. Therefore, if we count 71 and 58 operations per
interaction, theoretical peak FLOPS in our code with Intel Xeon
E5-2683 v4 is 50.6 and 54.5 GFLOPS, respectively.  From Figure
\ref{fig:ips}, the numbers of interactions from cells per second are
$\sim 7\times10^8$ at sufficiently large $N$.  For 71 and 58
operations per interaction, the measured performances of our code are
50 and 41 GFLOPS, which correspond to 99\% and 75\% of the peak.

\begin{figure}
\begin{center}
\includegraphics[width=8cm]{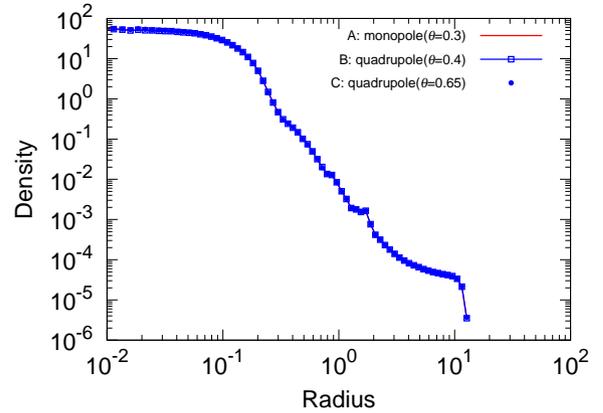}
\caption{
Radial density profiles of cold collapse simulations at $t=10$. Solid 
curves without points and with open squares show the results of 
simulations that utilized only monopole terms with $\theta=0.3$ (simulation A) and up 
to quadrupole terms with $\theta=0.4$ (simulation B), respectively. Circles show the 
result of simulation that utilized up to quadrupole terms with 
$\theta=0.65$ (simulation C).
}
\label{fig:rho}
\end{center}
\end{figure}

To validate effectiveness of our implementation for
astrophysical regimes, we performed three cold collapse simulations. We 
set the gravitational constant, the total mass of particles, the unit 
length, the total number of particles, the time step, and the softening 
length as $G=1$, $M=1$, $R=1$, $N=4,194,304$, $\Delta t=2^{-8}$, 
$\epsilon=2^{-8}$, respectively. The initial particle distribution was 
the homogeneous sphere whose radius is a unity, and the initial virial 
ratio was $0.1$. Three simulations were conducted on the machine shown in 
Table~\ref {tab:system} with 30 CPU cores. Differences between the three 
simulations are $\theta$ and whether quadrupole terms are calculated. 
Simulation A utilized only monopole terms with $\theta=0.3$. 
Simulation B and C calculated quadrupole terms and used $\theta=0.4$
and 0.65.
Figure~\ref{fig:rho} shows the radial density profiles of these simulations at $t=10$. 
Note that we plotted from $R=0.01$, which is about five times of 
$\epsilon=2^{-8}$.
The results of the three simulations agree well each other. 

The particle distribution is nearly homogeneous at $t<1$.
Thus, 
if we consider accuracy only at $t<1$, we can use $\theta=0.65$ when we calculate quadrupole terms to achieve comparable accuracy with only monopole terms as shown in Figure~\ref{fig:err_cc65k}.
The collapse occurs around $t=1$, and then
a dense flat core forms at $t>1$ as shown in Figure~\ref{fig:rho}.
Therefore, 
if we take account of accuracy at $t>1$, it is assumed that we should use $\theta=0.4$ rather than $\theta=0.65$ when quadrupole terms are adopted.
However, there
was little difference in the density profiles.  Thus, practically, we
might be able to use larger $\theta$ than expected to reduce the
calculation cost.
The average calculation time per step of these simulations
were 3.20 seconds, 3.70 seconds, and 2.05 seconds, 
for simulation A, B, and C, respectively. 
Therefore, we can gain 1.56 times better performance 
by calculating quadrupole terms with $\theta=0.65$ 
than by calculating only monopole terms of $\theta=0.3$.

As another practical astrophysical test, we performed a suite of cosmological
$N$-body simulations using the same initial condition.  The initial
condition consists of $128^3$ dark matter particles in a comoving box
of 103 Mpc and the mass resolution is $2.07 \times 10^{10} \,
M_{\odot}$.  We generated the initial condition at $z=33$ by a
publicly available code,
MUSIC \footnote{https://bitbucket.org/ohahn/music/} \citep{key-34}.
Here, we aim to evaluate the performance of our implementation for the
late phase of large scale structure formation.  For this reason, we
first simulated this initial condition down to $z=1$ by a TreePM code,
GreeM \citep{key-6, key-21}.  Then we identified particles within a
spherical region with a radius of 51 Mpc on the box center and added
hubble velocities to these particles.  We use these particles as the
new initial condition of our cosmological test calculations.

\begin{figure}
\begin{center}
\includegraphics[width=8cm]{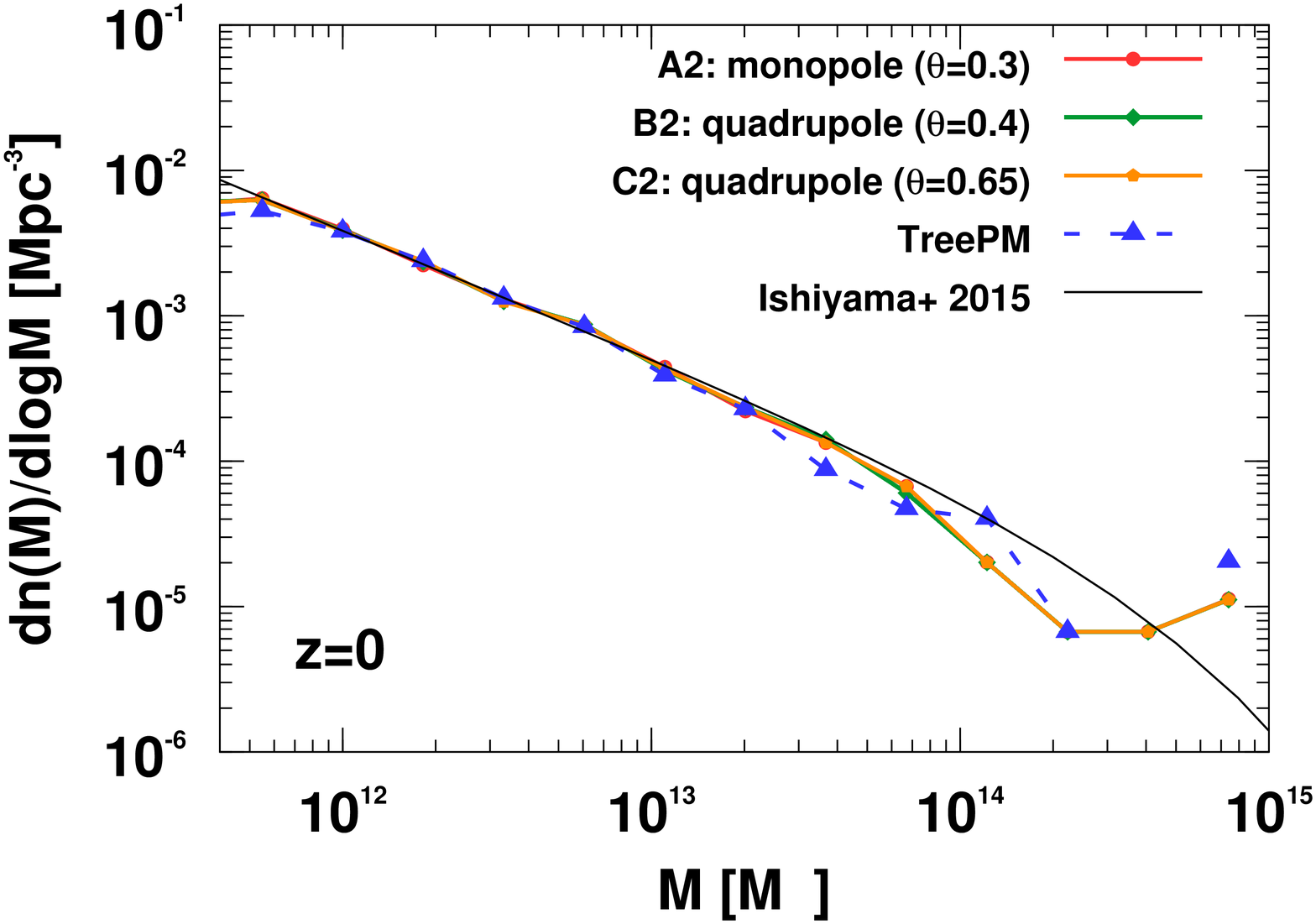}
\caption{
Mass functions of dark matter halos at $z=0$ obtained in a suite of cosmological test simulations.  Three solid curves with symbols are results from our implementation for three different settings.  Dashed curve is obtained from the simulation done by the TreePM method \citep{key-6, key-21}.  
Solid curve without symbols denotes a fitting function proposed by \citet{key-32}.
}
\label{fig:massfunc}
\end{center}
\end{figure}

We simulated the initial condition from $z=1$ to $z=0$ with three different
settings in the same manner as the cold collapse simulations shown
above, namely, simulation A2 which utilized only monopole terms with
$\theta=0.3$, simulation B2 and C2 which calculated quadrupole terms with
$\theta=0.4$ and $\theta=0.65$.
We also conducted the full box simulation by the
TreePM.  Figure~\ref{fig:massfunc} shows the mass functions of dark
matter halos at $z=0$, identified by ROCKSTAR phase space halo/subhalo
finder \citep{key-35}. The results of the three tree simulations and
TreePM simulation agree well each other and well fitted by a fitting
function calibrated by a suite of huge simulations \citep{key-32}.  

In the late phase of large scale strcuture formation such as $z<1$,
particle distributions are highly inhomogeneous because dense dark
matter halos form everywhere, indicating that it should be more
reasonable to use $\theta=0.4$ rather than $\theta=0.65$ when
quadrupole terms are adopted as discussed in cold collapse
simulations. However, the difference of halo mass functions is
indistinguishable.  Thus, practically, larger $\theta$ than expected
might be allowed to reduce the calculation cost.  The average
calculation time per step of these simulations were 0.415 seconds,
0.441 seconds, and 0.292 seconds, for simulation A2, B2, and C2, respectively.
Therefore, these results demonstrate that we can gain 1.42 times
better performance by our implementation.  These simple tests
reinforce the effectiveness of our implementation for some
astrophysical targets.

\section{Discussion}\label{sec:dis}
In this section, we estimate the performance of
our implementation on AVX-512 environment.
In AVX-512, the number of
SIMD registers is 32, which is twice of AVX2. This number is enough to hold data
that are currently needed to load every time the force calculation
loop is done. 
Line~18 to 23 in List~2 are the operations for loading coordinates of
$i$-particles.  Line~36 and 37 is the operation for loading the
gravitational softening length. Line~98 and 107 are the operations for
loading constant floating-point numbers, which are necessary to
calculate the quadrupole term of equation~(\ref{eq:pot}) and the
gravitational force given in equation~(\ref{eq:force}),
respectively. All data loaded by those operations do not change
throughout the entire $j$ loop in the force calculation from Line~16
to 123 in List~2. 
Therefore, Line~18 to 23,
Line~36, 37, 98, and 107 in List~2 can be moved to
before the loop.
Furthermore, the width of SIMD registers in AVX-512 is 512-bit, which
is twice of AVX2. 
This 
enables us to remove Line~56 in List~2 because the
elements of the quadrupole tensors of two $j$-cells, which are $6
\times 2=12$ elements, can be stored in one register. 
Without additional instructions, 
we can replace six VSHUFPS operations from Line~57 to 62 in List~2 to
six VPERMPS operations, which permute single-precision floating-point
value. The detail of VPERMPS is
available in IntelR 64 and IA-32 Architectures Software Developer’s
Manual~\footnote{https://software.intel.com/sites/default/files/managed/7c/f1/326018-sdm-vol-2c.pdf}.
Totally, we can
reduce the numbers of operations in the force loop from 59 to 48.
Furthermore, the AVX-512 instructions can simultaneously 
calculate 16 single-precision floating-point numbers because of the twice width of the SIMD registers.
Overall, we can estimate that the calculation of quadrupole terms
becomes $59/48\times2=2.46$ times faster in AVX-512 than AVX2. The
calculation of monopole terms will be twice faster in AVX-512 than
AVX2 because of the twice width of the SIMD registers.  It is
difficult to gain speed up in other parts such as the tree construction
and the tree traversal because hierarchical oct-tree structures are
used. 
Therefore, we assume that the calculation time for tree construction and tree traversal does not change on AVX-512 environment compared to that of AVX2 environment.

Table~\ref{tab:AVX-512} is the estimated ratios of the time for
calculating forces to that of using only the monopole on AVX-512
environment.  
Our implementation gives 1.08 times faster using the
quadrupole terms with $\theta=0.4$ than using only monopole terms with
$\theta=0.3$ for the Plummer model, and 1.02 times faster with
$\theta=0.65$ than that of $\theta=0.5$ with using monopole terms only
in the disk.

\begin{longtable}{cccc}
  \caption{Estimated ratios of the time for calculating forces and potentials to that of using only the monopole when we assume that the force
    calculation part is implemented with AVX-512. “Monopole”
    calculates only monopole terms. “Quadrupole” calculates
    quadrupole terms with our implementation. ``Homogeneous'' is the homogeneous sphere. ``Plummer'' is the Plummer model. ``Disk'' is the exponential disk. }\label{tab:AVX-512}
  \hline              
  Program & $\theta$ & Particle distribution & Ratio \\ 
\endfirsthead
  \hline
  Program & $\theta$ & Particle distribution & Ratio \\
\endhead
  \hline
\endfoot
  \hline
\endlastfoot
  \hline
  monopole & 0.3 & Homogeneous& 1\\
  quadrupole & 0.65 & Homogeneous& 0.44\\ 
  \hline
  monopole & 0.5 & Homogeneous& 1\\
  quadrupole & 0.75 & Homogeneous& 0.77\\ 
  \hline
  monopole & 0.3 & Plummer & 1 \\
  quadrupole & 0.4 & Plummer & 0.93\\ 
  \hline
  monopole & 0.5 & Plummer & 1\\
  quadrupole & 0.6 & Plummer & 1.26\\
  \hline
  monopole & 0.3 &Disk & 1\\
  quadrupole & 0.45 &Disk & 0.79\\ 
  \hline
  monopole & 0.5 &Disk & 1\\
  quadrupole & 0.65 &Disk & 0.98\\
\end{longtable}

\section{Summary}\label{sec:sum}
We have developed a highly-tuned software library to accelerate the
calculations of quadrupole term with the AVX2 instructions on the
basis of original Phantom-GRAPE (\cite{key-3}).  Our implementation
allows simulating homogeneous systems such as the large-scale
structure of the universe up to 2.2 times faster than that with only
monopole terms. Also, our implementation shows 1.1 times higher
performance than the combination of the pseudoparticle multipole
method and Phantom-GRAPE.  Further improvement of the performance is
estimated when we implement our code with the new SIMD instruction
set, AVX-512. On AVX-512 environment, our code is expected to be able
to accelerate simulations of clustered system up to 1.08 times faster
than that with only monopole terms.  Our implementation will be more
useful as the length of the SIMD registers gets longer. Our code in
this work will be publicly available at the official website of
Phantom-GRAPE~\footnote{https://bitbucket.org/kohji/phantom-grape}.

\begin{ack}
We thank the anonymous referee for his/her valuable comments.
We thank Kohji Yoshikawa, Ataru Tanikawa, and Takayuki Saitoh for
fruitful discussions and comments. This work has been supported by
MEXT as ``Priority Issue on Post-K computer'' (Elucidation of the
Fundamental Laws and Evolution of the Universe) and JICFuS. We thank
the support by MEXT/JSPS KAKENHI Grant Number 15H01030 and 17H04828.
This work was supported by the Chiba University SEEDS Fund (Chiba University Open Recruitment for International Exchange Program).
\end{ack}

\appendix

\end{document}